\documentclass[prb,twocolumn,showpacs,amsmath,amssymb,superscriptaddress]{revtex4}
\usepackage{graphicx}% Include figure files
\usepackage[english]{babel}
\usepackage{textcomp}
\begin{document}
\preprint{}
\title{A uniaxial stress capacitive dilatometer for high-resolution thermal expansion and magnetostriction under multiextreme conditions}
\author{R. K\"{u}chler}
%\email{kuechler@cpfs.mpg.de}
\affiliation{Max Planck Institute
for Chemical Physics of Solids, N\"{o}thnitzer Str. 40, 01187
Dresden, Germany}
\affiliation{Experimental Physics VI, Center for Electronic Correlations and Magnetism, University of Augsburg, Universit\"{a}tsstrasse 2, 86135 Augsburg, Germany}
\author{C. Stingl}
\author{P. Gegenwart}
\affiliation{Experimental Physics VI, Center for Electronic Correlations and Magnetism, University of Augsburg, Universit\"{a}tsstrasse 2, 86135 Augsburg, Germany}
\date{\today}
\begin{abstract}
Thermal expansion and magnetostriction are directional dependent thermodynamic quantities. For the characterization of novel quantum phases of matter it is required to study materials under multi-extreme conditions, in particular down to very low temperatures, in very high magnetic fields, as well as under high pressure. We developed a miniaturized capacitive dilatometer suitable for temperatures down to 20 mK and usage in high magnetic fields, which exerts a large spring force between 40 to 75 N on the sample. This corresponds to a uniaxial stress up to 3 kbar for a sample with cross-section of (0.5~mm)$^2$. We describe the design and performance test of the dilatometer which resolves length changes with high resolution of $0.02 \mathrm{\AA}$ at low temperatures. The miniaturized device can be utilized in any standard cryostat, including dilution refrigerators or the commercial physical property measurement system.
\end{abstract}

%\keywords{iron arsenide, high pressure, spin density wave, superconductivity}
\maketitle

\section{Introduction}

Thermal  expansion of solids is of great importance for technology and fundamental science. The linear coefficient $\alpha=L^{-1} (dL/dT)_p$ describes how the length of a piece of given material changes with temperature ($T$) at constant pressure ($p$). The volume expansion coefficient $\beta$, which for isotropic materials is given by $3 \cdot \alpha$ quantifies the relative change of the sample volume $V$ with temperature. Using Maxwell's relation the volume thermal expansion coefficient can be expressed as $\beta= -V^{-1} (dS/dp)_T$ where $S$ and $p$ denote entropy and pressure, respectively. A similar relation holds between the linear thermal expansion and the uniaxial pressure dependence of the entropy along the respective direction. Compared to specific heat, $C=T(dS/dT)_p$, the linear thermal expansion is a particularly interesting thermodynamic property, which provides complementary directional dependent information. Thermodynamic analysis allows to calculate the initial pressure dependence of a phase transition temperature from thermal expansion and specific heat. Again, respective information on the initial uniaxial pressure dependencies is obtained by analyzing the linear instead of the volume thermal expansion. Thus, thermal expansion is an important and sensitive tool to investigate all kind of phase transitions in condensed matter.

Thermal expansion is particularly interesting for studying low-temperature behavior of correlated electron systems [Ref.\cite{VisserA}]. Such systems are very susceptible to small external perturbation, which can lead to various kinds of phase transitions. Thermal expansion is a suitable tool for the detection of such transitions. As an example, the ground state of Kondo lattice materials sensitively depends on the balance of Kondo- and RKKY-interactions which both can be tuned by pressure. Such materials have a small characteristic energy scale $E^\ast$, which is highly pressure sensitive. This is highlighted by an enhanced Gr\"uneisen parameter $\Gamma = {(V_m E^\ast)^{-1}}   {\partial E^\ast / \partial p}$. In the experiment, different contributions from, e.g., phonons or electrons contribute to the thermal expansion and specific heat. It is convenient to calculate an "effective Gr\"uneisen parameter"
\begin{equation}
 \Gamma_{\rm eff}(T)=\frac{V_m}{\kappa_T} \frac{\beta(T)}{C(T)}=\sum \Gamma_i \frac{C_i(T)}{C(T)}
 \end{equation}
[Ref.\cite{VisserA}] which is the sum of the Gr\"uneisen parameters from the various contributions (e.g. phonons, electrons, magnons) times their relative fraction to the total heat capacity ($\kappa_T$ denotes the isothermal compressibility). $\Gamma_{\rm eff}$ is huge (of order 100) for Kondo lattice materials [Ref.\cite{VisserA,Kuechlerreview}]. Furthermore, for materials near a quantum critical point the characteristic energy scale $E^\ast$ is given by temperature itself in the quantum critical state. Consequently, the Gr\"uneisen ratio diverges as $T\rightarrow 0$. This makes thermal expansion an important tool for identifying quantum phase transitions [Ref.\cite{Kuechler1}]

So far we have discussed the application of thermal expansion for investigating the initial response to pressure in the limit of $p \rightarrow 0$. However, in many cases it is most interesting, to follow the behavior as a function of increasing pressure. Important examples include the study of superconductors (sometimes even superconductivity is only induced by pressure [Ref.\cite{Mathur}]) or magnets, which display changes of the ordering under pressure or even its suppression towards quantum critical points. Thermal expansion measurements under helium-gas pressure were already carried out many years ago [Ref.\cite{Fietz}] and up to 0.25~GPa previously [Ref.\cite{Manna}]. We focus our attention on uniaxial stress. Thermal expansion under (low) uniaxial stress (up to 100 bar) were already peformed by M. de Souza and L. Bartosch [Ref.\cite{Souza}]. For unconventional superconductors the application of sufficient uniaxial stress allows to obtain information on the order parameter symmetry. For example, a chiral $p_x\pm i p_y$ order parameter symmetry has been proposed in Sr$_2$RuO$_4$ and a symmetry breaking uniaxial pressure is expected to induce a splitting of the respective superconducting transitions [Ref.\cite{Hicks}].

Uniaxial stress experiments are particularly exciting for frustrated magnets. Currently, there is a growing interest in frustrated systems with many low-energy configurations which amplify quantum and thermal fluctuations in stabilizing novel (quantum) states of matter. Geometrical frustration describes situations where interactions are incompatible with the lattice geometry, e.g. the prevention of antiferromagnetic order between three spins located on the vertices of an equilateral triangle [Ref.\cite{Ramirez}]. Frustrated magnets have a huge (macroscopic) ground-state degeneracy [Ref.\cite{Moes}] which could enable spin liquid behavior. In systems with bond frustration like honeycomb A$_2$IrO$_3$ (A = Na,Li) [Ref.\cite{Singh}] with honeycomb Kitaev exchange, the application of uniaxial stress would also be very interesting, in order to tune the ground state from magnetic order into the Kitaev quantum spin liquid phase. In addition to frustrated insulating magnets, recently frustrated Kondo lattice materials have attracted considerable interest. Strong geometrical frustration counter acts the Kondo singlet formation and may enable a Kondo breakdown quantum critical point and metallic spin liquid state [Ref.\cite{Si,Senthil,Custers}]. It is thus highly desirable to access and modify geometrical frustration by application of uniaxial stress. 

For studying low-temperature excitations near quantum critical points (QCPs) or quantum spin liquid (QSL) phases, measurements require an extremely high sensitivity to detect a relative thermal length change of  $\Delta L/L < 10^{-9}$. The required resolution of $\Delta L = 10^{-2}$ $\mathrm{\AA}$ for samples with a length in the mm range can only be achieved by capacitive dilatometry [Ref.\cite{Manna,Kue,Pott,Braendli,Huy,Visser,Lorenz,Schmiedeshoff,Neumeier,Park,Steinitz}], which is a few orders of magnitude larger than the sensitivity of all other methods (optical methods:$\Delta L = 10^{0}$ $\mathrm{\AA}$  [Ref.\cite{Daou}], piezocantilever technique: $\Delta L = 10^{-1}$ $\mathrm{\AA}$ [Ref.\cite{Park}]).

A few years ago, we have developed one of today's best high-precision miniature capacitance dilatometers for measuring thermal expansion and magnetostriction [Ref.\cite{Kue}]. This is an extremely compact and miniaturized dilatometer constructed from a Be-Cu alloy using milling and spark erosion. The resolution of such a cell is determined by the diameter of the two capacitor plates and by the parallelism between them. The cells produced by our innovative, patent-pending production method are only marginally wider than the size of the capacitor plates [Ref.\cite{Kue}]. This optimized concept and the high level of manufacturing quality allow for an unprecedented resolution $\Delta L = 10^{-2} \mathrm{\AA}$  in a capacitive dilatometer of this compact size.

Modification of the design of [Ref.\cite{Kue}] allows to determine the length change with similar high resolution under a substantial uniaxial stress. This opens entirely new functionality for detecting pressure-induced phenomena. This paper introduces a uniaxial stress dilatometer. We also demonstrate the possibility to tune the degree of frustration in Kondo lattice materials. Here we take advantage of the fact that linear thermal expansion is a unidirectional thermodynamic probe, related to the uniaxial pressure derivative of entropy. It is highly sensitive to phase transitions as well as quantum criticality [Ref.\cite{Steppke,Kuechler1,Kuechler2,Kuechler3}] and therefore best suited to study the effect of tuning frustration by uniaxial stress.

In the next section, we first describe the design and construction of the uniaxial stress dilatometer in detail. The thermometry and the capacitance measurement setup are discussed only briefly since we used standard methods. The measuring process is described in chapter III. Here also the force applied to the sample by the two leaf springs is discussed and set in relation to the applied uniaxial stress. The dilatometer has been operated successfully in a PPMS (2\,--\,300 K, and in magnetic fields up to 10 T), in an exchange gas cryostat (3.5\,--\,300 K, and in fields up to 9 T), as well as in a dilution refrigerator with the dilatometer mounted in a vacuum (0.025\,--\,6~K in fields up to 20 T). We have applied a maximal uniaxial stress of 1 kbar. In Chapter IV, measurements of the thermal expansion between 2 and 300 K are shown. In IV.A, details of the cell operation and calibration, corrections due to the thermal expansion of the empty cell, and a thermal expansion test measurement of copper under uniaxial stress is presented. In IV.B, uniaxial stress thermal expansion on a polycrystalline Ni-Co-Mn-Sb Heusler alloy is shown. Finally, the functionality and the extremely high resolution of our uniaxial device is demonstrated by a measurement at millikelvin temperature on a CeRhSn single crystal (chapter V).

\section{THE UNIAXIAL STRESS DILATOMETER}
\begin{figure}[t]
\centering
\includegraphics[angle=0,width=5.3cm,clip]{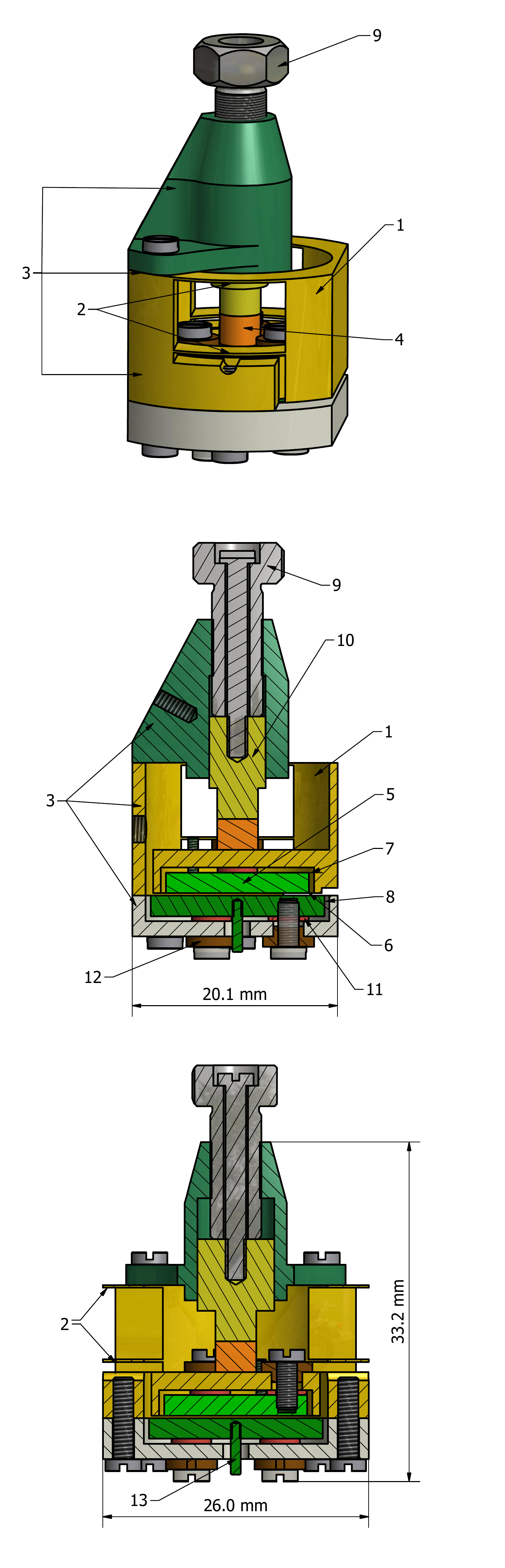}
\caption{\label{Fig1} Uniaxial stress capacitive dilatometer in three-dimensional (upper), side cut-away (middle) and front cut-away (lower) views (cf. for comparison the respective views on the original design in [Ref.\cite{Kue}]). The numbers illustrate: (1) mobile part, (2) 0.7 mm thick Be-Cu flat springs, (3) housing, (4) sample, (5) upper capacitor plate, (6) lower capacitor plate, (7) upper guard ring, (8) lower guard ring, (9) adjustment screw, (10) piston, (11) sapphire washer, (12) insulating piece of vespel, (13) electrical connection.}
\end{figure}

We have designed our new uniaxial stress dilatometer based on the patent-pending miniature dilatometer [Ref.\cite{Kue}] which operates on the principle of two parallel flat springs developed by Pott and Schefzyk [Ref.\cite{Pott}]. While the Pott-Schefzyk dilatometer was assembled from ten main parts which determined the dimension of the cell, our previous innovation in [Ref.\cite{Kue}] was to produce the corpus of the cell, which originally consisted out of six different parts, from a single piece of BeCu, using milling and spark erosion. New modifications allow to apply a substantial uniaxial stress.
A schematic of the uniaxial stress dilatometer is shown in Fig.~1. All parts (except some insulating spacers) were fabricated from high-purity beryllium copper to minimize Eddy current heating during magnetic field sweeps. The main body (yellow part of Fig.~1) contains the mobile part (1), both springs (2), as well as the middle part of the housing (3). While the lower capacitor plate (6) is part of the housing (3), the upper capacitor plate (5) is fixed to the mobile part (1), which is held in the frame by two very thick 0.7 mm BeCu leaf springs (2). The sample is held between the piston (10) and the mobile part including the upper capacitor plate (5) and can be tensioned by the adjustment screw (9). A length change of the sample (4) causes an equivalent displacement of the upper capacitor plate with respect to the lower one and induces a change of the capacitance. Samples of less than 1 mm and up 6 mm length can be measured. The two capacitor plates are electrically isolated by insulating pieces of vespel (12) and 0.5 mm sapphire washers (11) and are surrounded by guard rings (7, 8) to avoid stray electric fields. For both the lower (6) and the upper capacitor plate (5), three BeCu screws are used to mount the plates to the guard rings (7,8). Before assembling the dilatometer, the capacitor plates were polished within their frames. A uniform surface of the plates within their frames is most essential to achieve best parallel orientation of the plates. In its rest position, the capacitance of the dilatometer is about 7 pF, corresponding to a distance of 0.22 mm between the capacitor plates. After mounting the sample, the adjustment screw (9) is used to reduce this distance to about 0.07 mm, which corresponds to a capacitance of 20 pF. By careful construction of the capacitance measuring circuit (shielding, avoiding of ground loops etc.) the absolute value of the capacitance is measured by a commercial capacitance measuring bridge (Andeen Hagerling 2550A) with a resolution of 10$^{-6}$ pF, which corresponds to a relative sensitivity $\Delta L / L  \approx 10^{-9}$ for a sample length in between 1 and 6 mm. 

Our new design (see Fig.~1) allows measurements under substantial uniaxial stress. Compared to the original ``almost-zero pressure'' (AZP) cell, three fundamental changes were made. (a) Most important the thickness of the two leaf springs has been increased significantly from 0.25 mm to 0.7 mm. Our theoretical calculations showed that the springs force $F$, and accordingly the applied uniaxial pressure $p$, increase with growing spring thickness $d$ as $F \propto p \propto d^3$. Therefore we expect that the springs force to the sample will increase from 3 N up to about 70 N by using 0.7 mm instead of 0.25 mm thick springs. (b) The adjustment screw holder is bigger and more robust so that the entire construction can withstand the applied forces. (c) The cylindrical adjustment screw (9) contains an ovally shaped piston (10), which is fixed to the head of the screw. This prevents the piston from rotating during adjustment and thus avoids damage to the sample. Due to the substantial applied uniaxial stress, the adjustment screw can only be tightened by a wrench.

\section{SPRING FORCE EXERTED ON THE SAMPLE}

%A small spring force exerted on the sample by the two leaf springs is inevitable with the described design. This is important, since we intend to use this induced uniaxial stress to tune the sample properties.%
The new functionality is based on the enormous force exerted on the sample by the two springs. Previously, we have described in detail how a tiny spring force of 3 N, in our previous dilatometer with 0.25 mm thick leaf springs, has been determined [Ref.\cite{Kue}]. In most experiments, such a weak force has no influence on the material properties. However, by increasing the spring thickness, the spring force and as a consequence the induced uniaxial pressure (stress) grows strongly. In the following, we will determine the expected uniaxial stress for the new spring thickness of 0.7 mm.

\begin{figure}[t]
\centering
\includegraphics[angle=0,width=10cm,clip]{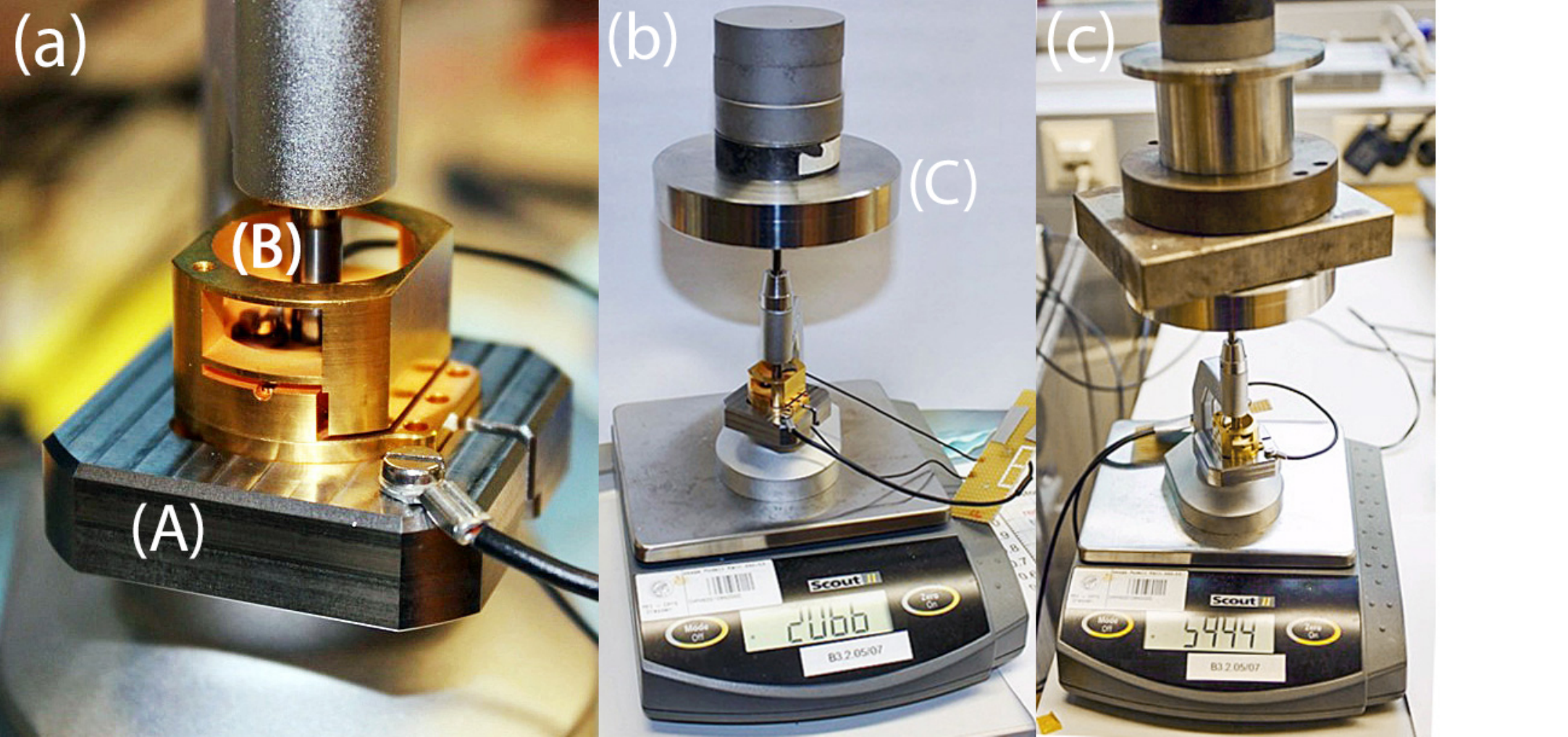}
\caption{\label{Fig2} Determination of the resulting spring force. (a) The tip of the pole (B) is set perpendicular to the upper surface of the movable part. (b) and (c) When the weight is increased up to 2066 g and 5994 g, respectively,  the pole tip increases its pressure on the movable part and causes a downward vertical movement of the movable part, including the upper capacitor plate.}
\end{figure}

\begin{figure}[t]
\centering

\includegraphics[angle=0,width=8cm,clip]{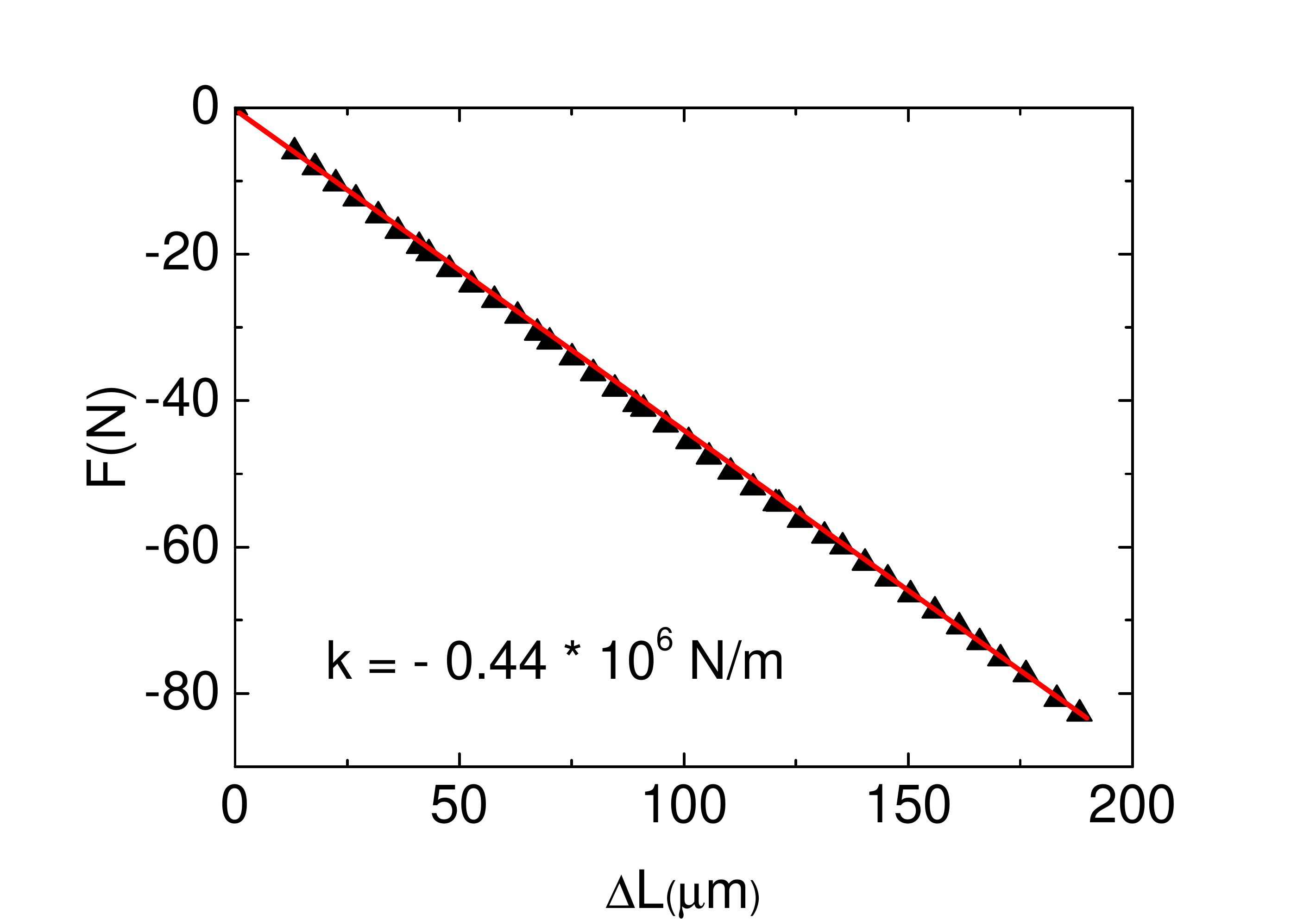}
\caption{\label{Fig3} Experimentally determined relation between the applied spring force and the displacement of the upper capacitor plate from its rest position, 220 $\mu$m above the lower plate. The red lines displays Hooke's law $F = k \times x$ with spring constant as quantified.}
\end{figure}

We use equation (3) to test the functionality of the new type of dilatometer at room temperature and to estimate the force exerted on the sample. Here, $\Delta L$ is the length change of the sample, i.e., the distance change between the capacitor plates. $C$ is the changing capacitance and $C_0$ the initial capacitance value. The main source of errors in the measurement are slightly non-parallel capacitor plates [Ref.\cite{Pott, Genossar}]. A measure of the tilt is given by the capacitance $C_\text{max}$, which is the maximum capacitance just before the capacitor shorts. We obtained the short-circuit capacitance by carefully decreasing the plate distance with the adjustment screw until the capacitor shorted. The final and highest value measured is $C_\text{max}$. The obtained value for $C_\text{max}$ = 150 pF, which confirms the high manufacturing quality. Taking into account the tilting of the plates, Pott and Schefzyk [Ref.\cite{Pott}] derived  equation (3), which is a corrected expression for the measured length change $\Delta L$, where $\epsilon_0=8.8542 \cdot 10^{-12}$ F/m is the permittivity in vacuum and $r = 7$ mm is the radius of the circularly shaped smaller upper capacitor plate.

\begin{equation}
\Delta L =  \epsilon_0\pi r^2   \frac{C-C_0}{C \cdot C_0} \left(1-\frac{C \cdot C_0}{C_\text{max}^2} \right).
\label{Funktionstest}
\end{equation}\\

Fig.~2 shows photos of the setup used to determine the applied spring force. A dilatometer holder plate (A) is mounted on a tripod foot. The tripod contains a guide slot in which a pole (B) is inserted. By placing weights on the disk (C), significant force can be applied via the pole to the mobile part of the cell. The setup is placed on a scale to measure the entire load.

%To estimate the spring force we placed little by little more and more weights on top of the disk. Hereby the tip of the pole is set perpendicular to the upper surface of the movable part. When the weight is increased, the pole tip increases its pressure on the movable part and causes a downward vertical movement of the movable part, including the upper capacitor plate. The change of capacitance is simultaneously measured by a commercial capacitance bridge. Since the whole setup is placed on a scale we could also measure the increasing weight simultaneously.

The mobile part of the cell was forced down stepwise with weights up to 8 kg. The resulting length change $\Delta L$ was calculated from the measured capacitance using equation (3). Fig.~3 shows the obtained linear relation between the length change and the applied force. From this we derive a spring constant $k = - 0.44 \times 10^{6}$ Nm$^{-1}$. This observation of a linear relation is very important as it proves the spring strain is still within the elastic regime for springs of 0.7 mm thickness. This means the uniaxial stress dilatometer can be used over the same range of operation as the AZP dilatometer. Since the elastic modulus of copper alloys remains almost constant from room temperature to below 1 K, this value is applicable to the entire temperature range. The determined relation between plate displacement, force and capacitance is shown in Fig. 4. As the dilatometer is operated between 15 and 40 pF, which corresponds to a plate distance of 100 - 25 $\mu$m, one obtains a spring force of 40 to 75 N. This corresponds to a maximal uniaxial stress of 0.75 and 3 kbar considering a cuboid sample with (1 mm)$^{2}$  and (0.5 mm)$^{2}$ cross-section, respectively. 

We also have manufactured and tested a dilatometer with 0.9 mm thick springs. It turned out that for this type the spring force is not constant in the entire range and Hooke's law is not obeyed. Thus our design only works for springs with a thickness less than 0.9 mm.

\begin{figure}[t]
\centering
\includegraphics[angle=0,width=8cm,clip]{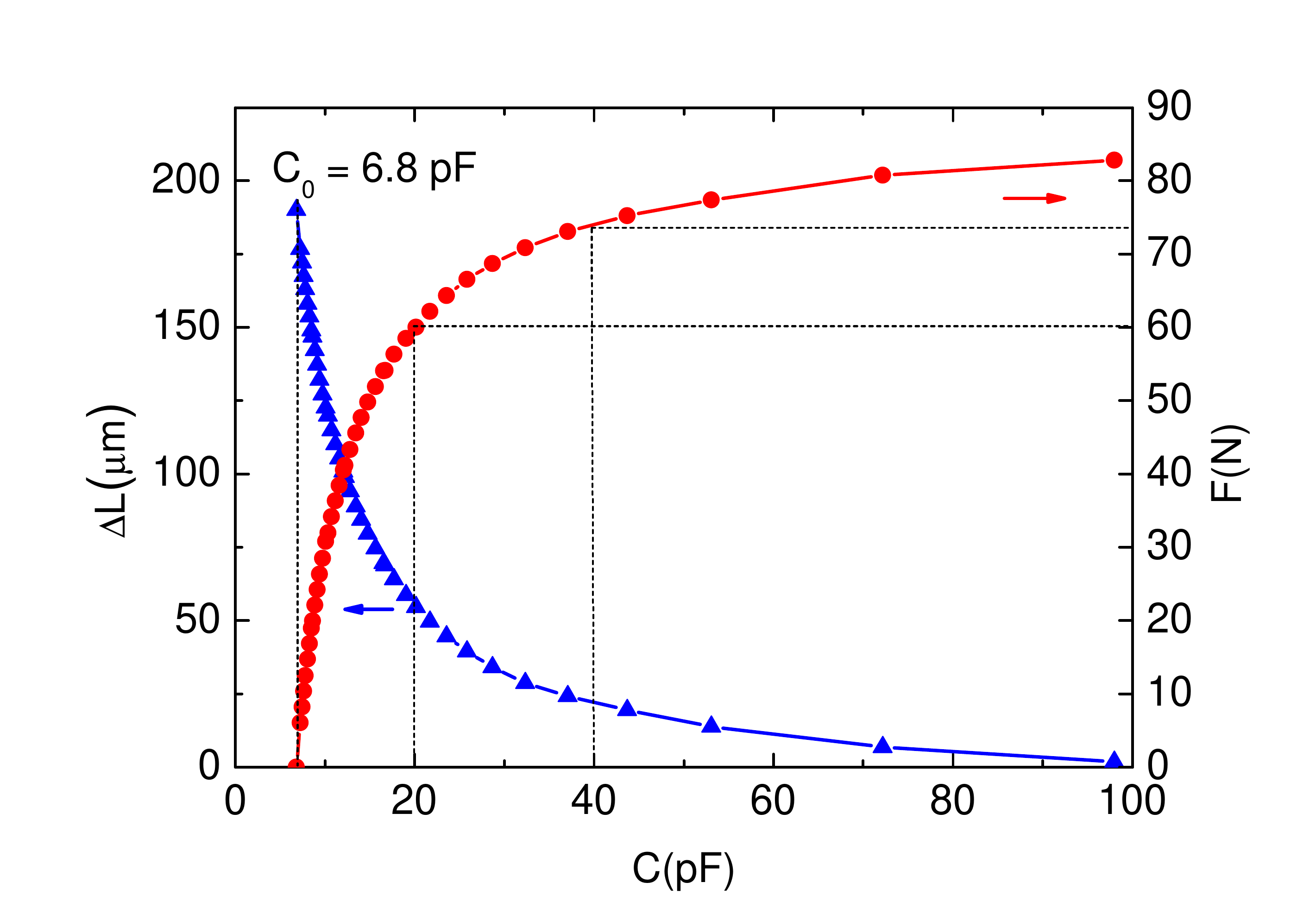}
\caption{\label{Fig4} Capacitor plate displacement and respective spring force as function of working capacitance. We typically operate the cell in between 20 and 40 pF corresponding to forces between 60 and 75 N indicated by dotted lines .}
\end{figure}

\section{APPLICATION TO THE PPMS}

\begin{figure}[t]
\centering
\includegraphics[angle=0,width=8cm,clip]{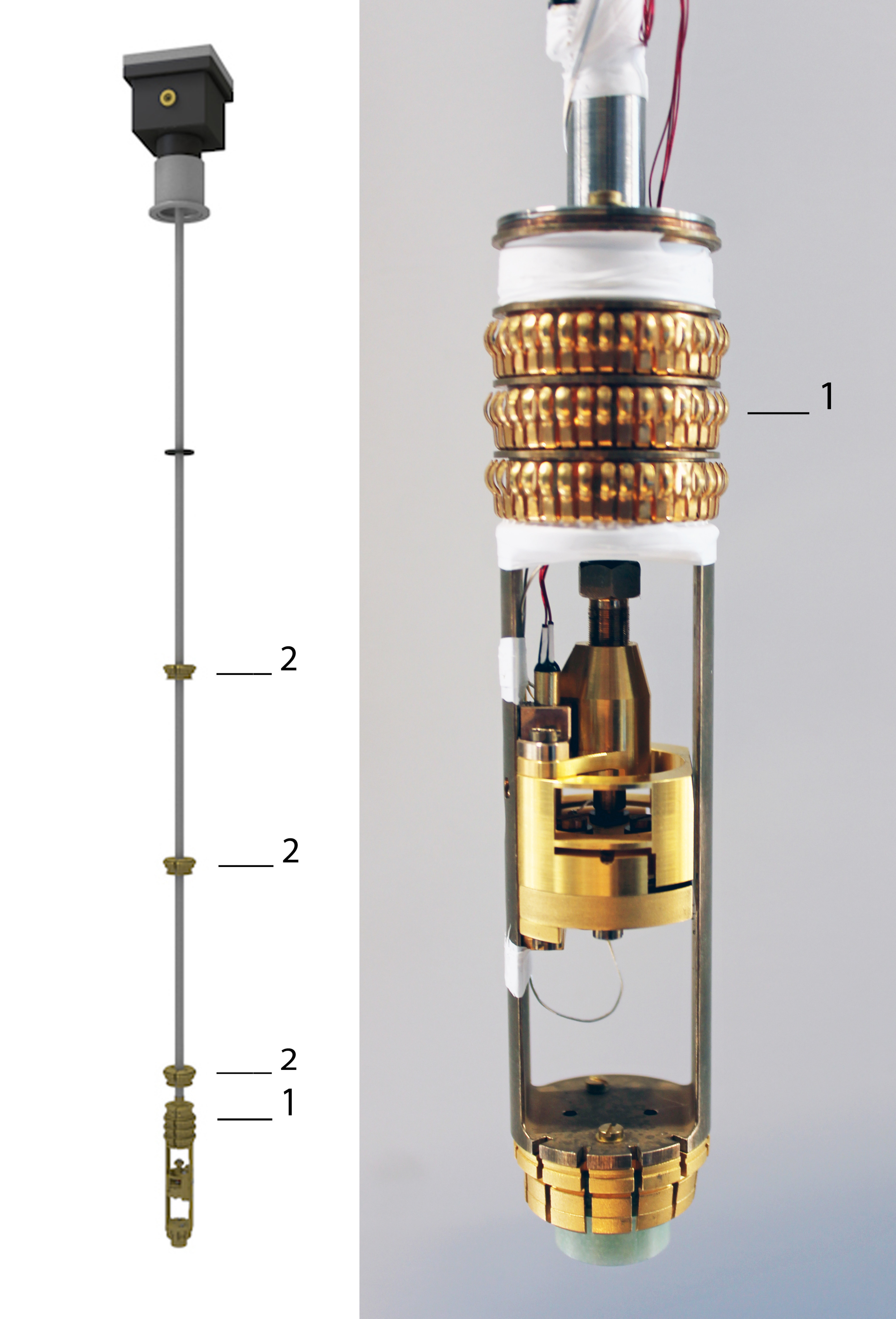}
\caption{\label{Fig6} Left: drawing of the PPMS probe with gold plated thermal anchors (1,2) and connectors for capacitance and temperature measurements. Right: photograph of uniaxial stress dilatometer within the PPMS probe.}
\end{figure}

Fig.~5 shows the PPMS uniaxial dilatometer probe. For operation between 2 and 300 K the sample space was kept under a helium atmosphere of typically 1 mbar, which ensured a good thermal contact of the cell and the included sample. Two coaxial cables are connecting the capacitor plates with the commercial measuring bridge. The PPMS insert is thermally coupled to the annulus via a pin connector, where a heater warms the helium gas to the correct temperature. Gold plated thermal anchors mounted to a copper block (1) just above the cell connect the probe to the walls of the IVC improving the thermal coupling to the cell. Additional anchors (2) are mounted above the insert at several points of the probe to reduce the temperature successively from the top of the cryostat down to the cell. The heat leak caused by the coaxial cables is reduced by wrapping them around the probe. A Cernox resistance thermometer is mounted close to the sample on the head of the cell.

\subsection{Cell background and performance test}

Similarly as for the AZP dilatometer [Ref.\cite{Kue}] we first need to determine a temperature-dependent background which arises from the thermal expansion of the different materials in the dilatometer assembly. The cell design, where nearly all components are machined from Be-Cu alloy, minimizes this background. Unavoidable exceptions are sapphire washers and electrically insulating parts made of vespel (cf. Fig.~1). We determined the cell background by a reference measurement of a copper sample. As performance test of the uniaxial stress dilatometer, we subsequently measured the thermal expansion of a piece of silver. The background subtracted results were compared to the literature. We measured 3 mm long samples of Cu and Ag with a cross-section of 3 mm diameter at a starting capacitance of 15 pF which corresponds to a uniaxial pressure $p_u$ of 71.4 bar ($p_u = F/A$ = 50 N/7 (mm)$^{2}$, the exerted force $F$ = 50 N corresponds to a measuring capacitance of 15 pF (see Fig.~4)) . 

During cooling from room temperature down to 2\,K the capacitance increases up to 15.8 pF. Thus the uniaxial pressure slightly changes up to 74 bar. We assume an applied uniaxial stress between 71 and 74 bar will not change the physical properties of copper and silver and as a consequence the measured thermal expansion should equal their thermal expansion behavior at AZP.     

Similar as previously [Ref.\cite{Kue}] we need to consider the changes of the sample length as well as the dilatometer cell upon changing temperature. The measured length change of the sample $\Delta L^\text{sample}_\text{meas}$ is the difference between the actual length change $\Delta L^\text{sample}$ of the sample and the length change of the cell $\Delta L^\text{cell}$:
\begin{equation}
    \Delta L^\text{sample}_\text{meas}=\Delta L^\text{sample}-\Delta L^\text{cell}
    \label{Zelleffektnicht}
\end{equation}\\
For calibration of the cell effect $\Delta L^\text{cell}$, we measured the thermal expansion of a piece of high-purity copper (99.999 \%) with known thermal expansion [Ref.\cite{Kroeger}]:

\begin{equation}
\Delta L_\text{meas}^\text{Cu} = \Delta L_\text{lit}^\text{Cu}-\Delta L^\text{cell}
\end{equation}\\
Combining (4) and (5) leads to
\begin{eqnarray}
&\Delta L^\text{sample} &=
{\Delta L}^\text{sample}_\text{meas}-{\Delta L}^\text{Cu}_\text{meas} + \Delta L_\text{lit}^\text{Cu}.
\end{eqnarray}
The relative length change of the sample normalized to its room temperature length $L_0$ results is
\begin{equation}
\left(\frac{\Delta L}{L_0}\right)^\text{sample} = \frac{\Delta L^\text{sample}_\text{meas}-{\Delta L}^\text{Cu}_\text{meas}}{L_0} +\left(\frac{\Delta L}{L}\right)^\text{Cu}_\text{lit} .
\end{equation}
Here the last term is the literature value for the relative thermal expansion of pure copper, which is independent of the sample length.

\begin{figure}[t]
\centering
\includegraphics[angle=0,width=9cm,clip]{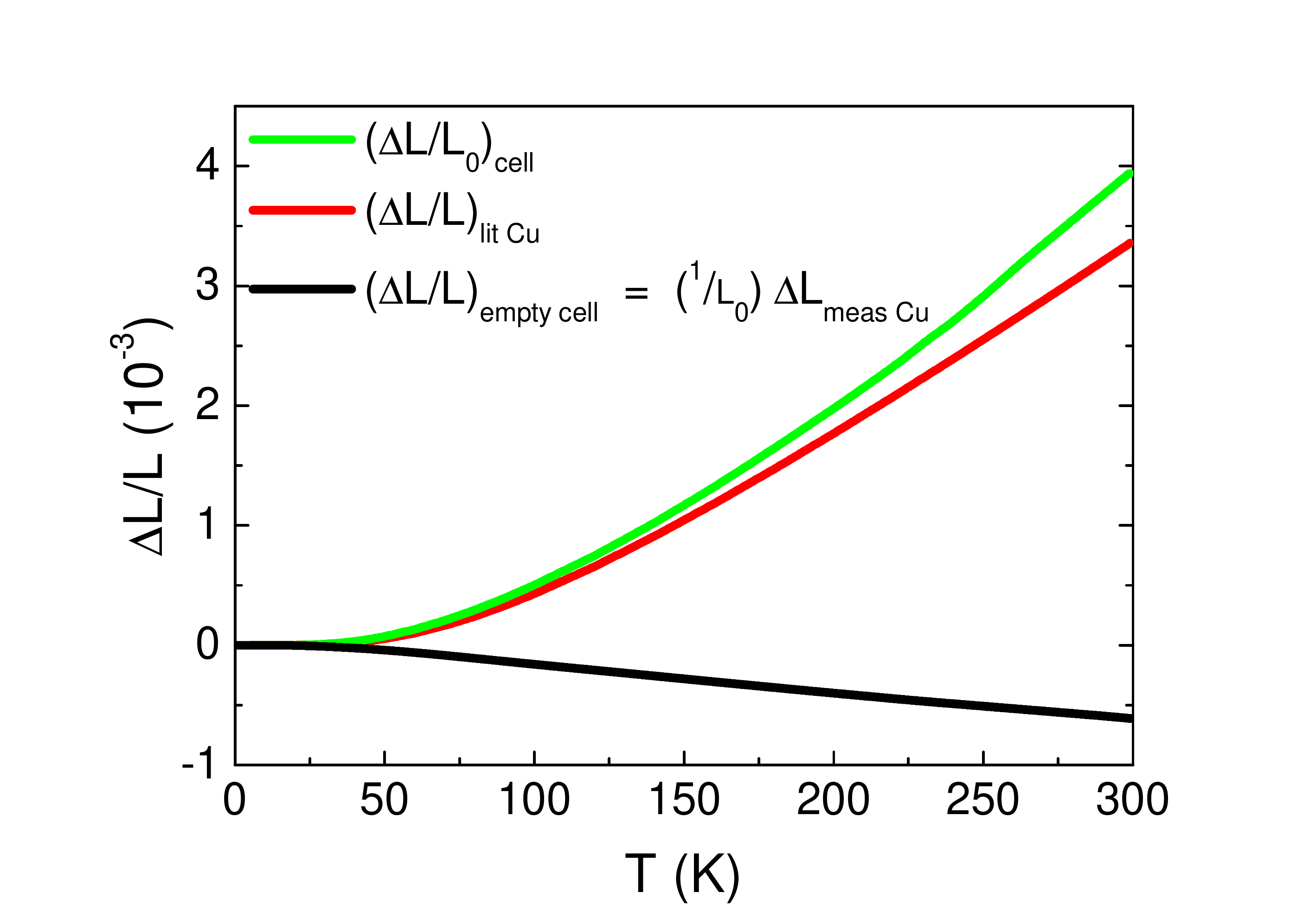}
\caption{\label{Fig7} Measured empty cell effect of the uniaxial stress dilatometer (black), literature values for the relative length change of copper (red) and calculated uniaxial stress cell effect (green).}
\end{figure}

Fig.~6 shows the copper measurement of the uniaxial stress dilatometer (black), the literature values of copper (red) and the resulting total cell effect (green) for a 3 mm long sample. Since the beryllium content $\mathrm{\textit{x} = 1.84}\,\%$ of the cell body material $\mathrm{Cu_{\textit{1-x}}Be_{\textit{x}}}$ is very low, the thermal expansion coefficient of the cell body material deviates only slightly from that of pure copper. Therefore the copper measurement resembles the empty cell effect: a measurement without sample where the adjustment screw is tightened to the bottom of the moveable part (see Fig.1). As one can see, the empty cell effect is quite small and shows nearly a linear temperature dependence. The thermal expansion of the cell deviates only slightly from a pure block of copper, demonstrating the high quality of the cell. The determined empty cell effect is found to be very similar to that of the AZP dilatometer [Ref.\cite{Kue}] as is expected from the similarities in the used components and main design.  

Above we described how we have obtained the cellbackground for a 3 mm large copper sample. In order to get the background for a sample of arbitrary length $L$ one has to consider the following. The different materials used for the insulation washers (vespel) and the capacitor gap (vacuum) cause a contribution to the thermal length change which is independent of the sample length. In contrast, thermal gradients introduce a background, which may have both sample length dependent and independent components. Thus, the cell background for arbitrary sample length $L$  is given by:         

\begin{equation}
    \Delta L^\text{cell} (T) = \Delta L^\text{cell}_{*} (T) +  \Delta L^\text{cell}_{**} (T) \cdot L ,
    \label{Zelleffektnicht}
\end{equation}    

where $\Delta L^\text{cell}_{*} (T)$ and $\Delta L^\text{cell}_{**} (T)$ denote the sample length independent and dependent contributions, respectively. Our numerous measurements of samples with different length have proven the validity of eq.(8), which showed that  $\Delta L^\text{cell} (T)$ is indeed a linear function of $L$. This makes it possible to obtain the cell background for a sample of arbitrary length just by interpolating between the two measured backgrounds for samples with smaller and larger length.

\begin{figure}[t]
\centering
\includegraphics[angle=0,width=8cm,clip]{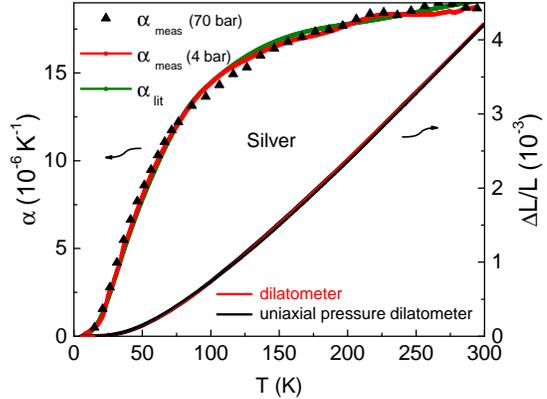}
\caption{\label{Fig8} right: Relative length change $\Delta L/L$ of silver under about 70 bar uniaxial stress (black) and comparison with literature data for $p_u = 0$ [Ref.\cite{Kroeger}] (red). left: thermal expansion coefficient $\alpha (T)$ of silver under about 70 bar uniaxial stress (black triangles) as function of temperature. The red curve shows the coefficient under 4 bar (measured with the AZP dilatometer [Ref.\cite{Kue}]) and the green curve the literature data for $p_u = 0$ [Ref.\cite{Kroeger}]}.
\end{figure}

To test the calibration of the cell effect and the sensitivity of the dilatometer, we performed an additional measurement of a silver sample (purity 99.999 \%, $l_{0}$ = 3 mm, 3 mm diameter) under the same pressure in between 71 and 74 bar. The results for the relative length change and the thermal expansion coefficient are shown in Fig.~7. The comparison with previous results for the AZP dilatometer [Ref.\cite{Kue}] as well as with literature results at ambient pressure ([Ref.\cite{White2, White3}]) reveals a very good agreement over the entire temperature range. For silver we do not expect a change of thermal expansion for uniaxial stress of order 100 bar. Thus, the result proves that the new device measures the thermal expansion coefficient with the same sensitivity as the AZP dilatometer. The uniaxial stress effect on physical properties will be demonstrated in the next sections.

\subsection{Effect of uniaxial stress on the thermal expansion of a polycrystalline Ni-Co-Mn-Sb Heusler alloy}

To demonstrate the functionality of our uniaxial stress dilatometer, we show the relative length change of a Ni$_{44}$Co$_6$Mn$_{38}$Sb$_{12}$ polycrystalline Heusler crystal with a length of $l = 2$ mm under AZP and 250 bar (see Fig.~8). The cell was mounted in a exchange gas cryostat but similar results can be obtained within a PPMS. At AZP and temperatures below $T_\text{C} =$ 351 K the Heusler alloy is a ferromagnet and undergoes a martensitic transformation at 217 K which drives the system into a weak-magnetic low-symmetry orthorhombic structure [Ref.\cite{Nayak}]. During cooling, a sharp step-like anomaly, observed in $\Delta L/L_0$ at $T_{M}$ = 217 K, indicates the first-order character of the phase transition. Hereby the crystal exhibits a contraction of about 0.5 \%. Upon subsequent heating, the length change is reversible within a thermal hysteresis of around 10 K. As can be seen in Fig.~8, the behavior of the thermal expansion under uniaxial stress is significantly changed. During cooling, the sample also exhibits a step-like jump, which is now broadened. The application of pressure shifts the martensitic transition to higher temperature as the pressure stabilizes the martensitic phase [Ref.\cite{Nayak2}]. The transformation from austenitic into martensitic variants already starts at $T =$ 235 K (M$_\text{s}$). The reason is that the low-symmetry orthorhombic structure of the martensitic phase has a smaller volume than the high-symmetry cubic structure of the austenitic phase [Ref.\cite{Nayak2}]. Also, the jump in the length change associated with the martensitic transformation is enhanced with increasing applied pressure. For AZP the relative length change at the martensitic transition is $\Delta L/L_0 \approx 0.5 \%$. Under a uniaxial stress of about 250 bar, the sample now shrinks almost five times stronger at the transition. This shows the effect of uniaxial stress on breaking the degeneracy associated with the symmetry-allowed martensitic variants. Hence, the increase of applied stress gives rise to a gradual increase in the fraction of martensitic variants which are energetically favorable to the direction of the applied uniaxial stress [Ref.\cite{Villa}]. It can also be seen that the transition under pressure is spread over a wider temperature range compared to ambient pressure. During heating the sample recovers to its initial shape not until room temperature. This irreversibility in cooling and heating could be related to different orientations of the martensitic variants or some kind of arrested effects in the sense that some fraction of the martensitic phase could not transform back into an austenitic phase before room temperature. It must be mentioned that, due to the extreme crystal contraction at the transition, the uniaxial stress applied on the sample is not constant during the temperature sweep. The pressure applied at room temperature (250 bar) decreases considerable down to 180 bar within the martensitic phase. This leads to a further artefactual broadening of the hysteresis. The transition starts in the austenitic phase (M$_\text{s}$) at the correct temperature but finishes the formation of the martensitic phase (M$_\text{f}$) at a lower temperature considering a uniform pressure of 250 bar. In this context it should be mentioned that such a strong sample contraction of about 3 \% is very exceptional and the here occurring strong change of the applied pressure will not take place in most applications.

\begin{figure}[t]
\centering
\includegraphics[angle=0,width=10cm,clip]{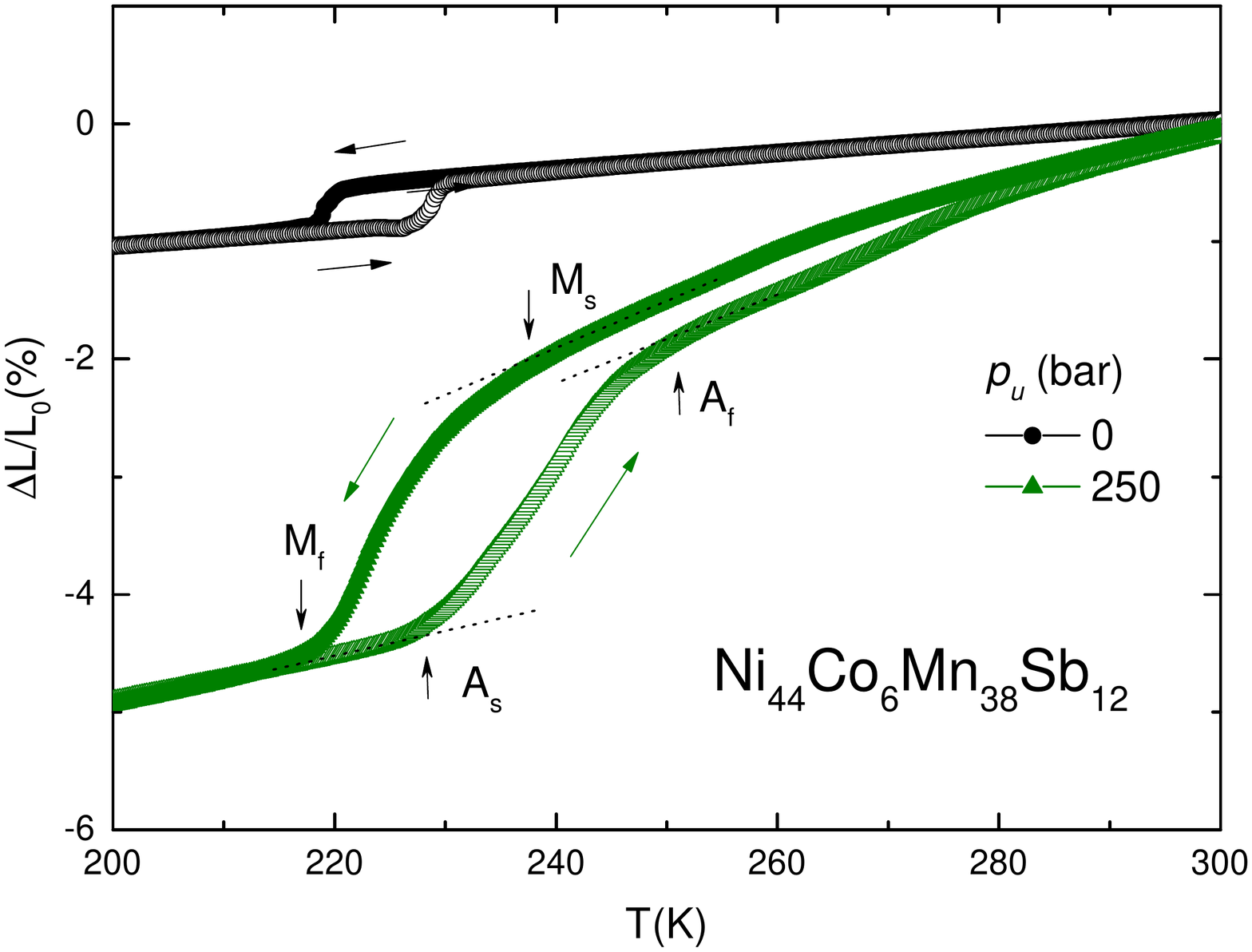}
\caption{\label{Fig8} Relative thermal length change of a Ni$_{44}$Co$_6$Mn$_{38}$Sb$_{12}$ polycrystalline Heusler crystal under ambient and 250 bar uniaxial stress. Closed and open symbols indicate measurements performed during cooling and heating, respectively.}
\end{figure}

\section{APPLICATION TO A DILUTION REFRIGERATOR}

\begin{figure}[t]
\centering
\includegraphics[angle=0,width=5cm,clip]{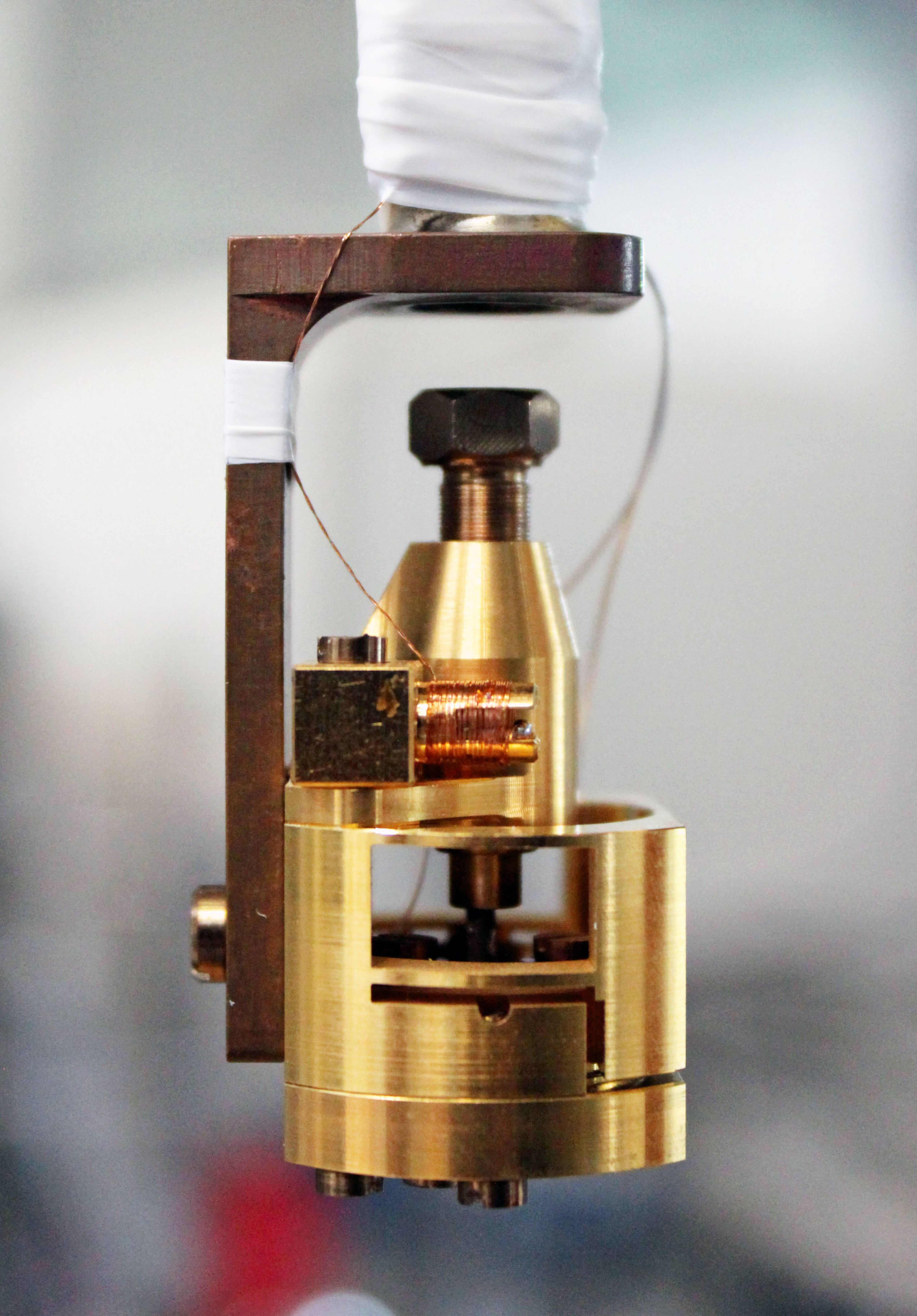}
\caption{\label{Fig10} Photograph of the uniaxial stress cell mounted on a cold finger of a dilution refrigerator.}
\end{figure}

\begin{figure}[t]
\centering
\includegraphics[angle=0,width=10cm,clip]{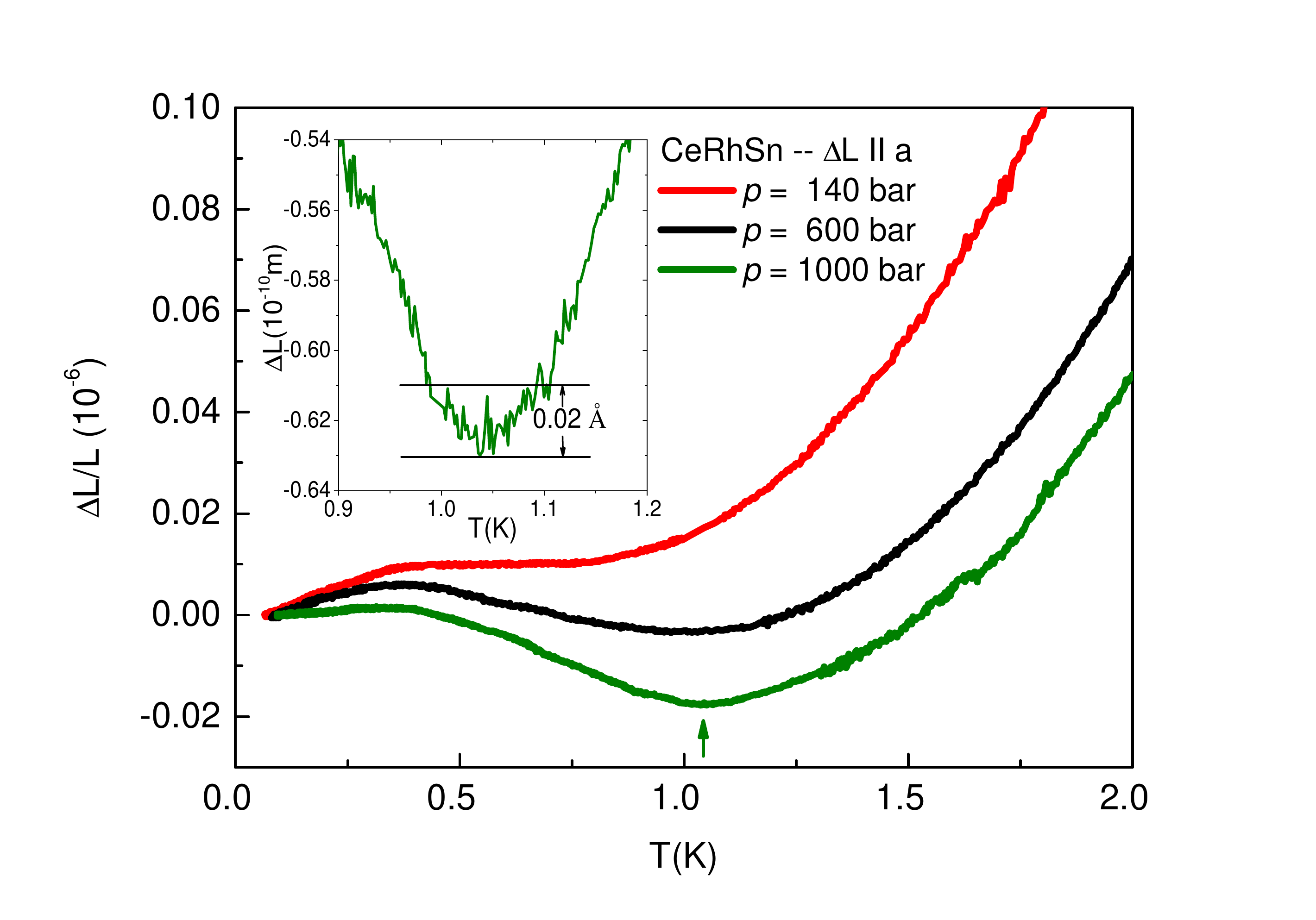}
\caption{\label{Fig9} Low-temperature relative thermal length changes $\Delta L(T)/T$ of a CeRhSn single crystal under uniaxial stress of 140, 600  and 1000 bar, measured along the \textit{a}-axis. The inset shows the high resolution of 0.02 $\mathrm{\AA}$ at low temperatures.}
\end{figure}

We demonstrate the functionality and sensitivity of our uniaxial stress dilatometer by low-temperature thermal expansion measurements of a CeRhSn single crystal with a length of $l = 3.5$ mm at various uniaxial stress up to 1000 bar. To perform this experiment the uniaxial stress cell is mounted on a cold finger of a dilution refrigerator as shown in Fig.~9. The hexagonal heavy fermion metal CeRhSn, with Ce atoms on distorted Kagome planes stacked along the c-axis, is an example for a geometrically frustrated Kondo lattice. Tokiwa et al. have studied the AZP linear thermal expansion and found a divergence of the (thermal) Gr\"{u}neisen ratio, as well as magnetic Gr\"{u}neisen parameter both evidencing zero-field and AZP quantum criticality [Ref.\cite{Tokiwa}]. Strikingly, the linear thermal expansion coefficient $\alpha(T)/T$ diverges upon cooling only along the in-plane \textit{a}-direction, but not along \textit{c}. Consequently, \textit{c}-axis uniaxial stress is not a relevant control parameter for this quantum criticality. By contrast, uniaxial stress parallel to the \textit{a} axis deforms the equilateral triangular units and thus reduces the geometrical frustration [Ref.\cite{Tokiwa}]. To demonstrate the effect of uniaxial stress, we have applied various uniaxial stress up to 1000 bar along this \textit{a}-axis which has the most prominent effect on geometrical frustration. Upon cooling from room temperature down to 50 mK, the capacitance changed only slightly from 16 pF to 15.9 pF. In the measured temperature range from 50 mK up to 6 K the capacitance of 15.9 pF first changed on the third decimal place. In this case the well defined and constant uniaxial pressure can be determined easily by reading the spring force - working capacitance curve (see Fig.~4) considering the square shaped sample surface $A$ of e.g. (0.75 mm)$^{2}$ ($p_u = F/A$ = 55 N/0.56 (mm)$^{2}$ $\approx$ 1000 bar). Fig.~10 shows the thermal length changes at 140, 600  and 1000 bar. As one can see, the gradient of the curve in the low temperature range below 0.3 K becomes shallower with increasing uniaxial stress and is almost zero for 1000 bar. This indicates that uniaxial stress parallel to the \textit{a} axis tunes the system away from quantum criticality. Even more remarkably the slope becomes negative for pressures above 600 bar at about 0.4 K. For the highest applied pressure of 1000 bar one can clearly observe a negative gradient followed by a smoothed kink at about 1.05 K (see green arrow). This suggests that uniaxial stress reduces the geometrical frustration very strongly and induces magnetic order out of the quantum critical spin liquid state. Our experiment proves that our uniaxial stress dilatometer is suited to study the effect of tuning frustration by uniaxial stress. The inset shows the extraordinary sensitivity of our dilatometer with a very high resolution of 0.02 $\mathrm{\AA}$. Magnetostriction data with the same sensitivity has been obtained as well. These results will be discussed elsewhere.

\section{CONCLUSION}

Based on the previous design of a compact high-resolution capacitive dilatometer [Ref.\cite{Kue}], we have developed a new device which provides a similarly high resolution for measurements under substantial uniaxial stress. In this article the layout, construction and calibration of the miniaturized uniaxial capacitance dilatometer for thermal expansion and magnetostriction measurements is described. Compared to the almost zero pressure dilatometer, the uniaxial stress dilatometer features much thicker springs, a much stronger adjustment screw holder as well as a special piston shape which avoids the action of shear forces on the sample during application of pressure. The uniaxial stress dilatometer has similar compact design which allows to mount it within the bore of typical superconducting magnets. Furthermore, it is easy to operate either in a PPMS or a dilution refrigerator and reaches a similarly high resolution of 0.02 $\mathrm{\AA}$ at low temperatures. The new functionality arises from the substantial spring force between 40 and 75 N acting on the sample. Depending on the surface area, for samples with two parallel surfaces, uniaxial stress up to 3 kbar for a cross-section of (0.5 mm)$^{2}$ can be applied. Uniaxial pressures of this magnitude are sufficient to induce drastic changes in physical behavior, as we have demonstrated by measurements of a Heusler allow as well as a geometrically frustrated magnet. Since this high-resolution uniaxial pressure dilatometer can be used in high fields and at low temperatures, various interesting investigations on quantum materials, which require such multi-extreme conditions, are now possible.

\section*{Acknowledgments}

We thank C. Salazar for performing the experiments on the Ni-Co-Mn-Sb Heusler alloy, A.K.Nayak and C. Felser for providing Ni-Co-Mn-Sb samples and M.S. Kim and T. Takabatake for providing  CeRhSn single crystals. Continued support by A.P. Mackenzie, M. Brando and M. Nicklas is acknowledged. We are also indebted to T. L\"{u}hmann for his strong efforts on the measurement software and R.S. Manna and F. Arnold for valuable comments on the manuscript. Appreciation is given to the members of our mechanical workshop J. Faltin, T. Thomas and J. Scharsach, who have manufactured all precision parts of the new dilatometer. This work is supported by the German Science Foundation through the projects KU 3287/1-1 and GE1640/8-1.

\end{document}